\documentclass[useAMS,usenatbib]{mn2e}
\usepackage{amsmath}
\usepackage{color}
\input{epsf}
\input{psfig.sty}

\newcommand{\aap}    {A\&A}
\newcommand{\apjs}   {ApJS}
\newcommand{\apj}    {ApJ}
\newcommand{\apjl}   {ApJL}

\newcommand{\mnras}  {MNRAS}

\newcommand{\physrep}{Phys. Rep.}
\newcommand{\nat}    {Nature}

\newcommand{\jcap}   {JCAP}

\title[Anisotropic Halo Model: Implementation and Numerical results]{Anisotropic Halo Model}
\author[Sgr\'o, Paz \& Merch\'an]
{ Mario A. Sgr\'o$^1$\thanks{E-mail: marioagustin@mail.oac.uncor.edu},
  Dante J. Paz$^1$  ,
  Manuel Merch\'an$^1$\\
 $^1$ Instituto de Astronom\'{\i}a Te\'orica y Experimental (UNC-CONICET), Observatorio Astron\'omico de C\'ordoba, \\
 Laprida 854, C\'ordoba, X500BGR, Argentina\\
}

\begin{document}
\date{Accepted 2 May 2013 . Received 8 February 2013}

\pagerange{\pageref{firstpage}--\pageref{lastpage}} \pubyear{2013}

\maketitle

\label{firstpage}

\begin{abstract}
In the present work, we extend the classic halo model for the large-scale matter distribution including a triaxial model
for the halo profiles and their alignments.  In particular, we derive general expressions for the halo-matter cross
correlation function.  In addition, by numerical integration, we obtain instances of the cross-correlation function
depending on the directions given by halo shape axes. These functions are called anisotropic cross-correlations. With
the aim of comparing our theoretical results with the simulations, we compute averaged anisotropic correlations in cones
with their symmetry axis along each shape direction of the centre halo. From these comparisons we characterise and
quantify the alignment of dark matter haloes on the $\Lambda$CDM context by means of the presented anisotropic halo
model. As our model requires multidimensional integral computation we implement a Monte Carlo method on GPU hardware
which allows us to increase the precision of the results whereas it improves the performance of the computation.
\end{abstract}


\section{Introduction}

In the mid-twentieth century, \citet{NeymanScott1952} published a work which would be the basis of the Halo Model. In
that paper the authors proposed a novel model for the spatial distribution of galaxies in the Universe. To build up
their model they imposed four main assumptions: (i) galaxies in the Universe lie inside galaxy clusters; (ii) the number
of galaxies belonging to a cluster obeys a probabilistic law; (iii) there is a probabilistic law which describes the
distribution of galaxies inside such galaxy clusters; and (iv) the cluster centres are placed in the space under a
quasi-uniform probabilistic law. Although this formalism was originally thought to describe the distribution of galaxies
it was later generalized to describe the dark matter distribution \citep[e.g.,][]{2000MNRAS.318..203S,
2000MNRAS.318.1144P, 2002Cooray_Halo}. 

In its standard form the model assumes that all the matter in the Universe reside in matter haloes which have a
spherical density profile. In addition the halo-halo self-correlation at large scales is governed by the linear power
spectrum. In this sense, the standard halo model does not include any assumption about alignments in the halo-halo
distribution. However, it is well known from numerical simulations, that haloes exhibit mildly aspherical shapes, with a
slight preference towards prolate forms (see for instance \citet{Paz2006} and references therein). Moreover, it has been
shown that their orientations are related to the surrounding structures such as filaments and large-scale walls
\citep{1996Natur.380..603B, 2005Colberg_Shape, 2006Altay_Shape, 2007Aragon_Spin, 2007Brunino_Shape, 2007Bett_Shape,
2005Kasun_Shape, 2006Allgood_Shape, 2006Basilakos_Shape,Hahn1,Cuesta1,2011Noh}. This results are also supported by
observations showing a good agreement with those obtained from the galaxy group shapes alignment with the galaxy
distribution \citep{Paz2006,2011Lau,Paz2011,2012Smargon}. Most of these studies perform statistics on the inclinations
of halo axes with respect to directions defined by the surrounding structure. This structure is characterized by the
particle distribution through various topological signatures such as filaments, walls or voids \citep[see for
instance][]{2013MNRAS.429.1286C}. While these studies agree about the presence of alignments within the $\Lambda$CDM
context, the magnitude of the alignment effect depends strongly on the geometrical definition of the surrounding
structure. On the other hand, \citet{PaperL} have introduce a new approach which is independent of any definition of
surrounding structure. This approach uses a modified version of the classical two-point cross-correlation function to
quantify the alignment between centre objects and the Large Scale Structure (LSS). Furthermore, this modified
cross-correlation function allows a robust comparison between numerical simulations and observational data. This
approach has successfully characterized alignments between galaxy shapes, group shapes and simulated dark matter haloes
with surrounding structure traced by galaxies or simulated structures \citep{PaperL, 2009Faltenbacher_corr, Paz2011,
2012Faltenbacher, 2012Schneider}.

Relevant to this work are the studies by \citet{SmithWatts}.  These authors investigated the effects of halo triaxiality
and LSS alignment on the isotropic power spectrum.  They found that the effect of halo triaxiality manifests a small
suppression of power at the 1-halo term, whereas in the 2-halo regime they found an upper limit of $10$\% of clustering
increment, by imposing perfectly aligned haloes. These correlations on halo triaxiality can affect cosmic shear studies
used to analyse weak lensing \citep{2007ApJ...655L...1B}. Moreover, an intrinsic alignment in the structure traced by
galaxies could result in a spurious contribution to the shear power spectrum \citep{2011MNRAS.415.1681H,
2010A&A...523A..60S, 2010A&A...523A...1J, 2010A&A...517A...4J, 2010MNRAS.402.2127S, 2007MNRAS.381.1197H,
2005A&A...441...47K, 2004MNRAS.347..895H, 2004ApJ...601L...1T, 2003MNRAS.339..711H, 2003A&A...398...23K,
2002MNRAS.335L..89J, 2002MNRAS.333..501B}.

In this paper we derive an analytical model for the anisotropic halo-matter cross correlation function. We implement an
extension of the classical halo model for the large-scale matter distribution, that includes triaxial modelling of halo
profiles \citep[following][]{JingSuto} and their alignments (described in \S \ref{sec:Anisotropic-Halo-Model}). The use
of the correlation function instead of the power spectrum allows a straightforward implementation of alignment effects
and triaxiality of the haloes. Furthermore, this anisotropic cross-correlation functions can be easily measured from
observations and numerical results as showed in previous works \citep{PaperL,Paz2011}. Therefore, the main goal of the
present work is to obtain a suitable analytic model for these functions in order to compare with numerical.

The organization of this paper is as follow. In \S \ref{sec:Anisotropic-Halo-Model} we introduce the anisotropic halo
model and we describe some assumptions performed. In \S \ref{sec:Implementation} we describe the numerical
implementation. In \S \ref{sec:Comparison} we compare the halo-matter cross-correlation functions obtained from our
model with those estimated from N-body simulations. Finally, we discuss and summarize our results in \S
\ref{sec:Conclusions}.

\section[]{Anisotropic Halo Model}
\label{sec:Anisotropic-Halo-Model}

In this section we introduce the analytic development of a new halo model which takes into account the triaxial nature
of the haloes and its alignment with the environment. As in the standard halo model, we will assume that the matter
distribution is made up of distinct haloes in a wide range of masses. We characterize each dark matter halo in terms of
its mass normalized density profile $U(\vec{r},m,\vec{a},\vec{\epsilon}) \equiv \rho(\vec{r},m,\vec{a},\vec{\epsilon}) /
m$ where $\vec{r}$ is the position respect to the centre of an halo of mass $m$, the components of $\vec{a}$ are the
eigenvalues of the shape tensor of the halo (with $a_{1} > a_{2} > a_{3}$), and $\vec{\epsilon}$ indicates the direction
of the eigenvectors (i.e. its components are the tree Euler angles of the shape ellipsoid orientation). Assuming that
all dark matter lies inside haloes, the density field $\rho(\vec{x})$ can be computed as
\begin{equation}
 \rho(\vec{x}) = \sum_{i=1}^{N} m_{i} U(\vec{x}-\vec{x}_{i},m_{i},\vec{a}_{i},\vec{\epsilon}_{i})
 \label{rho}
\end{equation}
where $N$ is the total number of haloes in the Universe.  Similarly, the number density of haloes $n_c(\vec{x})$ can be
expressed as follow
\begin{equation}
 n_c(\vec{x}) = \sum_{i=1}^{N} \delta(\vec{x}-\vec{x}_{i})
\psi(m_{i},\vec{a}_{i},\vec{\epsilon}_{i})
 \label{nc}
\end{equation}
where the factor $\psi(m,\vec{a},\vec{\epsilon})$ is the selection function, and $\delta(\vec{x}-\vec{x}_{i})$ is the
Dirac delta function centred at the halo position.

The probability distribution $p(m,\vec{a},\vec{\epsilon})$ of haloes with mass $m$, shape vector $\vec{a}$ and
orientation $\vec{\epsilon}$ can be computed using the equation
\begin{equation}
 p(m,\vec{a},\vec{\epsilon}) = \left\langle \sum_{i=1}^{N} \delta(m-m_{i})
\delta( \vec{a} - \vec{a}_{i} ) \delta( \vec{\epsilon} - \vec{\epsilon}_{i} )
\delta( \vec{x} - \vec{x}_{i} ) \right\rangle\,,
\label{massfuncgen}
\end{equation}
where the angle brackets denote the average over an ensemble of realizations. In case that shapes and orientations are
not taken into account, this function reduces to the standard mass function. From the above definitions, the mean dark
matter density and the mean number density can be computed as:
\begin{equation}
 \overline{\rho} = \int dm \, d\vec{a} \, d\vec{\epsilon} \, m p(m,\vec{a},\vec{\epsilon})
\end{equation}
\begin{equation}
 \overline{n_{c}} = \int dm \, d\vec{a}  \, d\vec{\epsilon} \, p(m,\vec{a},\vec{\epsilon})
\psi(m,\vec{a},\vec{\epsilon})
\end{equation}

Hereafter the integration limits should be understood as taken over the whole range where the integrand is valid. The
$\Psi(m,\vec a, \vec \epsilon)$ function allows us to impose some restrictions on the mass range, shape and orientation
of the centre haloes.

Using these function we are able to compute the halo-matter cross correlation function as follow:
\begin{eqnarray}
 \xi_{hm}(\vec{r}) & = & \left\langle \left(\frac{n_{c}(\vec{x})}{\overline{n_{c}}} - 1\right)
                   \left( \frac{\rho(\vec{x}+\vec{r})}{\overline{\rho}} -1 \right) \right\rangle
                   \nonumber \\
                   & = & \left\langle
                   \frac{n_{c}(\vec{x})\rho(\vec{x}+\vec{r})}{\overline{n_{c}}\quad\overline{\rho}}
- \frac{n_{c}(\vec{x})}{\overline{n_{c}}} - \frac{\rho(\vec{x}+\vec{r})}{\overline{\rho}} + 1
\right\rangle \nonumber \\
                   & = & \left\langle
\frac{n_{c}(\vec{x})\rho(\vec{x}+\vec{r})}{\overline{n_{c}} \quad \overline{\rho}}
\right\rangle -1
\label{xihm}
\end{eqnarray}
where we have assumed the ensemble average can be replaced by an space average:
\begin{equation}
 \left\langle \frac{\rho(\vec{x}+\vec{r})}{\overline{\rho}} \right\rangle =
\frac{1}{V\overline{\rho}} \int d\vec{x} \, \rho(\vec{x+\vec{r}}) = 1
\end{equation}
and similarly
\begin{equation}
 \left\langle \frac{n_{c}(\vec{x})}{\overline{n_{c}}} \right\rangle =
\frac{1}{V\overline{n_{c}}} \int d\vec{x} \, n_{c}(\vec{x}) = 1
\end{equation}

Replacing \ref{rho} and \ref{nc} in equation \ref{xihm} we obtain:
\begin{align}
\xi_{hm}(\vec{r}) = -1 &+ \frac{1}{\overline{\rho}\overline{n_{c}}}\left\langle
      \left(\sum_{i=1}^{N} \delta(\vec{x}-\vec{x}_{i}) \psi(m_{i},\vec{a}_{i},\vec{\epsilon}_{i})\right)\right.
      \nonumber\\
      &\times\left.\left( \sum_{j=1}^{N} m_{j} U(\vec{x}+\vec{r}-\vec{x}_{j},m_{j},\vec{a}_{j},\vec{\epsilon}_{j})
\right)\right\rangle
\end{align}

It is possible to split this equation in two terms, those where $i=j$ (hereafter 1-halo term), and
those where $i \ne j$ (hereafter 2-halo term). Consequently we define:
\begin{align}
\xi_{hm}^{1h}(\vec{r}) = \frac{1}{\overline{\rho}\overline{n_{c}}} \Bigg\langle 
&\sum_{i=1}^{N} \delta(\vec{x}-\vec{x}_{i}) \psi(m_{i},\vec{a}_{i},\vec{\epsilon}_{i})\nonumber\\
            &\times m_{i} U(\vec{x}+\vec{r}-\vec{x}_{i},m_{i},\vec{a}_{i},\vec{\epsilon}_{i})\Bigg\rangle
\label{xi1h}
\end{align}
\begin{align}
\xi_{hm}^{2h}(\vec{r}) = &-1 + \frac{1}{\overline{\rho}\overline{n_{c}}} \left\langle 
\left( \sum_{i=1}^{N} \delta(\vec{x}-\vec{x}_{i}) \psi(m_{i},\vec{a}_{i},\vec{\epsilon}_{i}) \right)\right.\nonumber \\
&\times \Bigg( \sum^{N}_{ \substack{j=0\\j \neq i} } m_{j}
U(\vec{x}+\vec{r}-\vec{x}_{j},m_{j},\vec{a}_{j},\vec{\epsilon}_{j}) \Bigg)
\Bigg\rangle
\label{xi2h}
\end{align}

Rewritting equation \ref{xi1h} as:
\begin{eqnarray}
\xi_{hm}^{1h}(\vec{r}) &=& \frac{1}{\overline{\rho}\overline{n_{c}}} \frac{1}{V}
\sum_{i=1}^{N} m_{i} \psi(m_i,\vec{a}_{i},\vec{\epsilon_{i}}) \nonumber\\
&& \times\int d\vec{x} \,
\delta(\vec{x}-\vec{x}_{i}) 
U(\vec{x}+\vec{r}-\vec{x}_i,m_{i},\vec{a}_{i},\vec{\epsilon}_{i})   \nonumber \\
&=& \frac{1}{\overline{\rho}\overline{n_{c}}} \frac{1}{V}
\sum_{i=1}^{N} m_{i} \psi(m_i,\vec{a}_{i},\vec{\epsilon_{i}})
U(\vec{r},m_{i},\vec{a}_{i},\vec{\epsilon}_{i})  \nonumber \\
& = & \frac{1}{\overline{\rho}\overline{n_{c}}} \int dm \,
      d\vec{a} \, d\vec{\epsilon} \, m \psi(m,\vec{a},\vec{\epsilon})
      U(\vec{r},m,\vec{a},\vec{\epsilon}) \nonumber \\ 
&&    \times \frac{1}{V} \sum_{i=1}^{N} \delta(m-m_{i}) \delta(\vec{a}-\vec{a}_{i})
      \delta(\vec{\epsilon}-\vec{\epsilon}_{i}) \nonumber
\end{eqnarray}

Using equation \ref{massfuncgen}, the final expression for the 1-halo term results:
\begin{align}
 \xi_{hm}^{1h}(\vec{r}) = \frac{1}{\overline{\rho}\overline{n_{c}}} & \int dm \,
      d\vec{a} \, d\vec{\epsilon} \, m U(\vec{r},m,\vec{a},\vec{\epsilon})\nonumber \\
      &\times p(m,\vec{a},\vec{\epsilon}) \psi(m,\vec{a},\vec{\epsilon}) .
 \label{xi1hf}
\end{align}

A similar procedure can be used to derive the 2-halo term from equation \ref{xi2h}:
\begin{align}
\xi_{hm}^{2h}(\vec{r}) &= 
-1 + \frac{1}{\overline{\rho}\overline{n_{c}}} \int dm_{1} \, d\vec{a}_1\,
d\vec{\epsilon}_{1} \, dm_{2} \, d\vec{a}_{2} \, d\vec{\epsilon}_{2} \, d\vec{y} \,\nonumber \\
&\times m_{2}U(\vec{r}-\vec{y},m_{2},\vec{a}_{2},\vec{\epsilon}_{2})\psi(m_{1},\vec{a}_{1},\vec{\epsilon}_{1})\nonumber \\
&\times\frac{1}{V} \sum_{ \substack{j=0\\j \neq i} }^{N} \sum_{i=1}^{N} 
\delta(m_{1}-m_{i}) \delta(\vec{a}_{1}-\vec{a}_{i}) \delta(\vec{\epsilon}_{1}-\vec{\epsilon}_{i})\nonumber\\
&\times\delta(m_{2}-m_{j}) \delta(\vec{a}_{2}-\vec{a}_{j}) \delta(\vec{\epsilon}_{2}-\vec{\epsilon}_{j})
\delta(\vec{y}-(\vec{x}_{j} - \vec{x}_{i}))
\end{align}

The joint probability to have a pair of haloes with a given set of properties (mass, shape and orientation) at a
distance $|\vec{y}|$, can be written
\begin{align}
\frac{1}{V} \sum_{ \substack{j=0\\j \neq i} }^{N} \sum_{i=1}^{N} 
&\times\delta(m_{1}-m_{i}) \delta(\vec{a}_{1}-\vec{a}_{i}) \delta(\vec{\epsilon}_{1}-\vec{\epsilon}_{i})\nonumber\\
&\times\delta(m_{2}-m_{j}) \delta(\vec{a}_{2}-\vec{a}_{j}) \delta(\vec{\epsilon}_{2}-\vec{\epsilon}_{j})\nonumber \\
&\times\delta(\vec{y}-(\vec{x}_{j} - \vec{x}_{i}))={p}_{1} {p}_{2} (\xi_{1,2}(\vec{y})+1) ,
\end{align}
where $p_i=p(m_{i},\vec{a}_{i},\vec{\epsilon}_{i})$, $i=1,2$, are the probabilities to have a halo with a given set of
properties (mass, shape and orientation), and
$\xi_{1,2}(\vec{y})+1=\xi(\vec{y}_{1,2},m_{1},\vec{a}_{1},\vec{\epsilon}_{1},m_{2},\vec{a}_{2},\vec{\epsilon}_ {2})+1$
is the join probability to have a pair of haloes with the given properties separated by a distance $\vec{y}_{1,2}$. In
the right hand side we have assumed that the probability of halo properties ($p_1$ and $p_2$) and the join probability
to have a pair of haloes separated by a distance $\vec{y}$ are independent. Therefore, using this last equation, the
final expression for the 2-halo term is given by:
\begin{align}
  \xi_{hm}^{2h}(\vec{r}) = \frac{1}{\overline{\rho}\overline{n_{c}}} 
  &\int dm_{1}\, d\vec{a}_{1}\, d\vec{\epsilon}_{1}\, dm_{2}\, d\vec{a}_{2}\, d\vec{\epsilon}_{2}\,
  d\vec{y} \nonumber \\
  &\times m_{2}  U(\vec{r}-\vec{y},m_{2},\vec{a}_{2},\vec{\epsilon}_{2}) p_1 p_2
  \xi_{1,2}(\vec{y})\nonumber \\
  &\times \psi(m_{1},\vec{a}_{1},\vec{\epsilon}_{1})
  \label{xi2hf}
\end{align}

Unlike standard models, equations \ref{xi1hf} and \ref{xi2hf} give a halo model for the two-point anisotropic
cross-correlation function taken into account the shape and the orientation of haloes. We will devote the next two
sections to explain our implementation of this model in order to compare with the corresponding measured correlations in
the simulation.

\subsection{The 1-halo term}
\label{sec:1halo}

In order to estimate the 1-halo term (equation \ref{xi1hf}), we need to adopt models for the halo normalized density
profile ($U$) and for the distribution of halo parameters ($p$). Regarding the density profile, we use the triaxial
model described by \citet{JingSuto} (hereafter JS). This model assumes that isodensity surfaces of dark matter 	can be
described by triaxial ellipsoids with a shape vector $\vec{a}$ and an orientation $\vec{\epsilon}$. A simple way to
parametrize those surfaces is through the radial parameter $R$ as defined by \citet{SmithWatts}:
\begin{equation} 
\frac{R^2}{{a_3}^2} = \frac{z^2}{{a_1}^2} + \frac{y^2}{{a_2}^2} + \frac{x^2}{{a_3}^2}
\label{isodef}\ .
\end{equation}
Accordingly to JS, the former definition allows to express the halo density profile similarly to that in \citet{NFW}
(NFW profile): 
\begin{equation}
\frac{\rho(R)}{\rho_{\rm crit}}=\frac{\delta_{\rm c}}
{\left(R/R_{\rm 0}\right)\left(1+R/R_{\rm 0}\right)^{2}},
\label{nfw}
\end{equation}
where $\delta_{\rm c}$ is a characteristic density and $R_{\rm 0}$ is a scale radius. Analogously to the scale radius in
NFW profile, JS define $R_{\rm 0}= c_e/r_e$, where $c_e$ is the concentration parameter and $r_e$ is the characteristic
radius. JS also provides the relations $c_e=\chi c_{\rm vir}$, $r_e=\chi r_{\rm vir}$, where $c_{\rm vir}$, $r_{\rm
vir}$ are the virial concentration and radius, respectively. It is possible to relate the characteristic density
$\delta_{\rm c}$ with the concentration parameter by requiring that the mean density inside an ellipsoid with major axis
$r_e$ to be $\Delta_e$ times the critical density, obtaining:
\begin{equation}
  \delta_{\rm c} = \frac{\Delta_e}{3} \frac{c_e^3}{ln(1+c_e)-c_e/(1+c_e)}
\end{equation}
JS provides a relation involving this the equivalent density $\Delta_e$ as function of the virialization density in a
spherical NFW:
\begin{equation}
  \Delta_e=5\Delta_{\rm vir} \left(\frac{a_1^2}{a_3 a_2}\right)^{0.75}.
\end{equation}
A value of $\chi=0.45$ was obtained by the authors fitting numerical simulations. We found that a $\chi$ value of $0.5$,
is also consistent with simulation data and provides a better agreement between the NFW profile and the JS profile at
the spherical shape limit. In this limit case, our choice for the $\chi$ parameter gives similar values for the
$\delta_{\rm c}$ and the characteristic density of the NFW profile in a mass range of $10^8$ to $10^{15}$ $M_\odot$.
  
Once the normalized density profile has been established, we must set up the probability distribution
$p(\vec{a},m,\vec{\epsilon})$. For simplicity, we assume that the orientation $\vec{\epsilon}$ and the mass $m$ of a
given halo are independent random variables, whereas the halo shape vector $\vec{a}$ is only dependent on the mass.
Consequently the probability distribution can be written as:
\begin{equation}
p(m,\vec{\epsilon},\vec{a}) = n(m) p(\vec{\epsilon}) p(\vec{a}|m)
\end{equation}
where $n(m)$ is the halo mass function, $p(\vec{\epsilon})$ and $p(\vec{a}|m)$ are the orientation and shape probability
distributions, respectively. For instance, in case of a uniform  probability for the halo orientation on the sphere,
the density function $p(\vec{\epsilon})$ takes the form:
\begin{eqnarray} 
p(\vec{\epsilon})d\vec{\epsilon} &\equiv& p(\alpha,\beta,\gamma) d\alpha d\beta d\gamma \\ \nonumber 
                                      &=& \frac{1}{2\pi} \frac{1}{2} \frac{1}{2\pi}d\alpha d(cos\beta) d\gamma 
\end{eqnarray}
where the Euler angles are restricted to the ranges: $0 \ge \alpha \ge 2\pi$, $0 \ge \beta \ge \pi$, and $0 \ge \gamma
\ge 2\pi$.\\

Given that equation \ref{isodef} can be written in term of the ratios $a_{21}=a_2/a_1$ and $a_{32}=a_3/a_2$, we define
the $p(\vec{a}|m)$ conditional probability distribution as a function of these quotients $p(a_{21},a_{32}|m)$, where
both arguments are restricted to the interval $(0,\,1)$. With this simplification, we are able to exchange $d\vec{a} \to
da_{21}da_{32}$ in equations \ref{xi1hf} and \ref{xi2hf}. We approximate this distribution by a product of two Gaussian
laws and quadratic factors that ensure null probability in the interval limits \footnote{These factors ensure that a
perfect sphere ($a_{21} = a_{32} = 1$) or an ellipsoid with a null semi-axis have a null measurement}. Consequently, the
shape distribution takes the following form:
\begin{align}
 p(a_{21},a_{32}|m) \propto & a_{21}\,(1-a_{21})\,e^{-\frac{(a_{21} - A)^2}{2\sigma^2}} \nonumber \\ 
                     \times & a_{32}\,(1-a_{32})\,e^{-\frac{(a_{32} - B)^2}{2\sigma^2}}
\label{eq:forma}
\end{align}
where $A$ and $B$ are two mass dependent parameters which must be estimated, whereas a fixed value $\sigma = 0.1$ is
adopted. A theoretical approach to this function was made by \citet{LeeJingSuto}. However, their results are not in good
agreement with the numerical simulation. On the other hand, JS used a direct measurement of the axis ratios on dark
matter haloes extracted from N-body simulations with the aim to find the density distribution. In our work, we adopt the
analytical form given by equation \ref{eq:forma}, which resembles the shape distribution obtained from the simulations.
The two free parameters of the adopted distribution are used to fit the cross-correlation function.

The last ingredient required in order to estimate the 1-halo term is the mass function $n(m)$, which describes the
abundance of dark matter haloes in a mass interval around $m$. This function is estimated using the analytical formulae
provided by \citet{ShethMoTormen2001}.

\subsection{The 2-halo term}
\label{sec:2halo}

In the current section we will set the corresponding functions for the 2-halo term. To compute equation \ref{xi2hf} we
need to specify the function $\xi_{1,2}(\vec{y}) \equiv
\xi(\vec{y}_{1,2}|m_1,\vec{a}_1,\vec{\epsilon}_1,m_2,\vec{a}_2,\vec{\epsilon} _2)$. On large scales the relation between
haloes and matter overdensity is deterministic, therefore, as in the standard halo model, we can approximate $\xi_{1,2}$
by the following equation:
\begin{equation}
\xi(\vec{y})_{1,2} \approx
f(\vec{y}_{1,2}|\vec{a}_1,\vec{\epsilon}_1,\vec{a}_2,\vec{\epsilon}_2)b(m_1)b(m_2)\xi_{lin}(\vec{y}
_{ 1,2}) 
\end{equation}
where $b(m_1)$ and $b(m_2)$ are the bias factors for the centre and tracer haloes, respectively, and $\xi_{lin}(\vec{y}
_{ 1,2})$ is the Fourier Transform of the Linear Power Spectrum (LPS). We have explicitly introduced the factor
$f(\vec{y}_{1,2}|\vec{a}_1,\vec{\epsilon}_1,\vec{a}_2,\vec{\epsilon}_2)$ to take into account the different alignments.
We have adopted the analytical expression of the bias function given by \citet{ShethMoTormen2001} whereas the LPS is
computed using the approach given by \citet{EisensteinHu99}.

\begin{figure} 
  \epsfxsize=0.5\textwidth
  \centerline{\epsffile{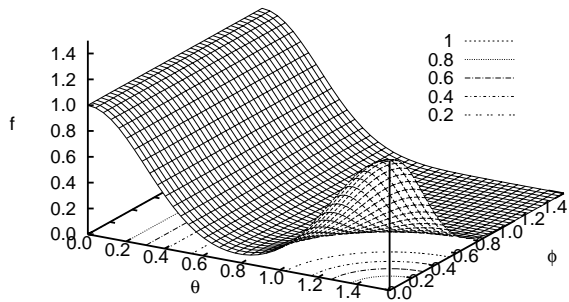}}
  \caption
  {
    Normalized alignment density distribution around dark matter haloes with parameters $C = 0.3$ and $D = 1.0$ (see
    equation \ref{eq:align}).
  }
  \label{fig:sabana}
\end{figure}

The $f$ function defined above can be used to describe two types of alignments: (i) the alignment between the position
of tracer haloes and the major-axis direction of centre halo, (ii) the alignment between the orientations of centre and
tracer haloes. It is reasonable to expect that the last kind of alignment has a second order effect on the model. This
can be explained by the fact that our model, and previous works \citep[see for
instance][]{2012vanDaalen,SmithWattsSheth2006,SmithWatts}, are insensitive to the halo shapes in the 2-halo term. This
is expected given that the difference between the major and minor axes of a halo is small compared to the distances, at
these scales, between centre and tracer haloes. Consequently, for the sake of  simplicity, we have assumed that the
alignment function does not depend on the alignment between the major axes of centre and tracer haloes. Furthermore, we
will consider this function as independent of halo shapes. Based on the previous considerations we adopt the following
expression:
\begin{equation}
f(\vec{y}_{1,2}|\vec{\epsilon}_1) \propto e^{-\frac{\theta^2}{2\,C^2}} + D\,e^{-\frac{(\theta - \pi/2)^2}{2\,C^2}}\,e^{-\frac{\phi^2}{2\,C^2}}
\label{eq:align}
\end{equation}
where $\theta$ and $\phi$ are the polar and azimuth angle of $\vec{y}_{1,2}$ in a spherical coordinate system defined by
the directions given by $\vec{\epsilon}_1$, respectively. The first (second) term in the preceding equation can be
understood as the excess of correlation along the major (intermediate) axis. As the alignment function is a probability
density distribution, the constant of proportionality is defined by its normalization. The $C$ and $D$ factors are free
parameters that we will determine in \S \ref{MLM}. Figure \ref{fig:sabana} shows an illustrative example of the
alignment function with parameters $C = 0.3$ and $D = 1.0$. It should be noticed that even when $\theta$ approach to $0$
and $\phi$ tend to be undetermined, the alignment nicely becomes constant by construction, avoiding any undesirable
behaviour on its evaluation. The increment of the function towards low $\theta$ values represents the alignment of the
matter distribution around the major shape axis, whereas the increment towards $\theta=\pi/2$ and $\phi=0$ represents
the corresponding alignment with the intermediate axis.

\begin{figure*}
  \epsfxsize=1.0\textwidth
  \epsffile{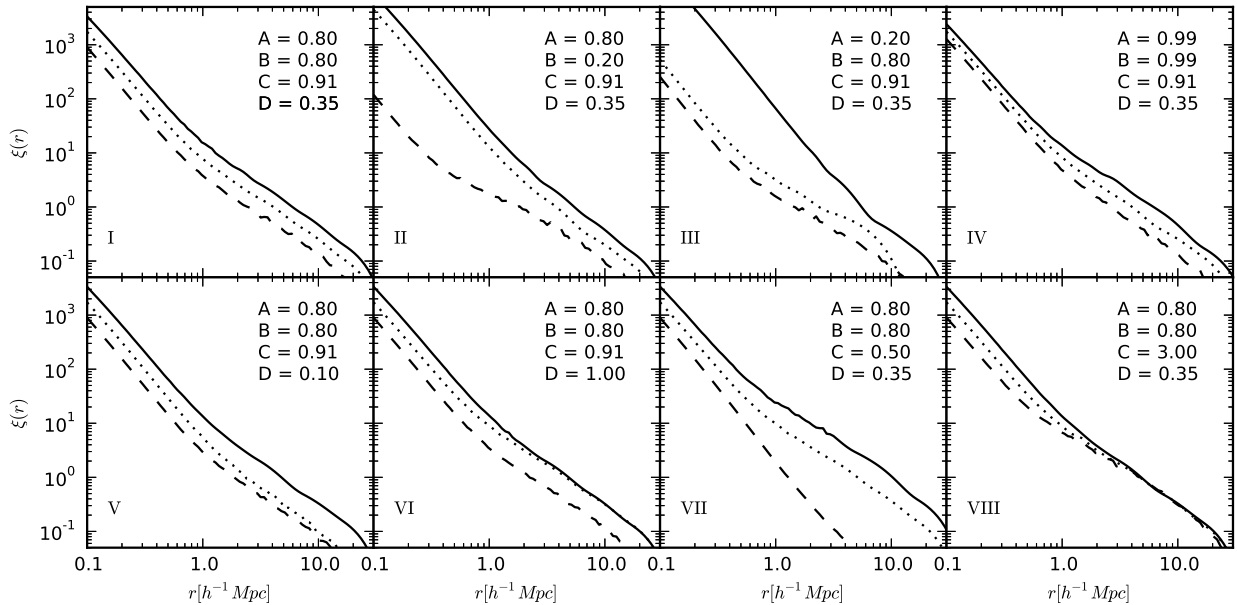}
  \caption
  {Results of the anisotropic cross-correlation functions along the halo shape axis directions. Centre halo masses span
a range from $10^{12}$ to $10^{16} M_{\odot} h^{-1}$.Solid, dotted and dashed lines correspond to major, intermediate
and minor axis direction, respectively. Each panel corresponds to a different set of parameters as explained in the
text.}
  \label{fig:direction}
\end{figure*}

\section{Implementation}
\label{sec:Implementation}
In this section we first describe the numerical method used to compute the integrals \ref{xi1hf} and \ref{xi2hf}.
Secondly, we show the results for the three-dimensional anisotropic cross-correlation function.

\subsection{Integration method}

Given the high order integrations involved in the 1-halo and 2-halo terms of our model, standard numerical methods (e.g.
quadrature rules) are not feasible to implement. In order to compute the multidimensional integrals we have employed a
Monte Carlo technique (MC). According to this method, the $m-$dimensional integral $I$ of a given function $f(\vec{x})$
in the volume $V \subset R^m$  can be approximated by means of the following expression:
\begin{equation}
 I = \int_{V} f(\vec{x}) dx_1dx_2...dx_m \approx \frac{V_{c}}{N} \sum_{i=1}^N
f(\vec{x}_i)
\label{IMC}
\end{equation}
where $\vec{x}_i = (x_i^{1},..,x_i^{m})$, are $N$ uniform random vectors within the computational volume $V_{c}
\supseteq V$. It can be shown that the uncertainty in the determination of $I$ decreases as $N^{-1/2}$ independently of
the dimension of the integrand. It is possible to improve the precision of the MC integration by using the importance
sampling method. This method requires the specification of a density function $w(\vec{x})$ for the random variable
$\vec{x}$. Thus, in the right hand side of equation \ref{IMC}, $f(\vec{x}_i)$ must be replaced by
$f(\vec{x}_i)/w(\vec{x}_i)$. It is worth to mention that using standard integration methods, for a fixed number of
subdivisions $N$ of the volume $V$, the error $\sigma_I$ quickly increases as the number of dimensions increase
($\sigma_I \propto N^{-1/m}$).\\

As stated in the preceding paragraph, to estimate the integrals \ref{xi1hf} and \ref{xi2hf} we should generate as many
random numbers as possible. The Monte Carlo integration is an embarrassingly parallel problem, so we have employed the
Nvidia CUDA extension for C language in order to compute the integrals. This choice allows us to increase the amount of
random numbers and consequently to enhance the precision of the results while the total running time grows moderately.

\subsection{The anisotropic tridimensional cross-correlation function}

Before comparing our halo model with results obtained from numerical simulations, we compute the cross-correlation
function along the shape axis directions of the centre halo in order to give a qualitatively analysis of the parameter
dependence. For this purpose, we fix the orientation of the centre halo such that the directions of the shape axes
correspond to cartesian axes (denoted by $\hat{e_1},\,\hat{e_2},\,\hat{e_3}$, respectively). Accordingly to this choice,
the integral in equation \ref{xi1hf} (\ref{xi2hf}) is reduced to a 3-D (12-D) integration. We define three correlation
functions by taking $\vec{r} \parallel \hat{e_i}$ where $i=1,\, 2,\, 3$. 

With the aim of describing qualitatively the behaviour of our model for different values of $A$, $B$, $C$ \& $D$, figure
\ref{fig:direction} shows the anisotropic cross-correlations for haloes in the mass range $10^{12} - 10^{16} M_{\odot}
h^{-1}$. The three directions, major, intermediate and minor axis are shown with solid, dotted and dashed lines,
respectively. Each panel corresponds to a different set of parameters. Upper panels show the correlation function
behaviour when the shape parameters $A$ and $B$ are changed. As can be seen in panel $(\mathrm{I})$ when $A = B$ and $C$
\& $D$ are properly set, the three functions show roughly constant differences. Panel $\mathrm{II}$ ($\mathrm{III}$)
shows the results for prolate (oblate) halo shapes $B < A$ ($A < B$) whereas panel $\mathrm{IV}$ shows the behaviour for
a nearly spherical halo shape, i.e. $A = B \approx 1$. Regarding to the alignment function, lower panels in figure
\ref{fig:direction} illustrate the effect of changing $C$ and $D$ parameters. Panels $\mathrm{V}$ and $\mathrm{VI}$
($\mathrm{VII}$ and $\mathrm{VIII}$) show the influence when assuming low and high values of $D$ ($C$), respectively. As
can be appreciated, for large values of the $C$ parameter the alignment vanishes because the alignment distribution
function is nearly constant. From this qualitative analysis, it can be seen that the behaviour of the model shows a
large versatility with the parameters variation. Furthermore, as expected, the parameters $A$ \& $B$ mainly affect the
1-halo term whereas the parameters $C$ \& $D$ are involved only in the 2-halo computation.

\section[]{Comparison with Numerical Simulations}
\label{sec:Comparison}

\begin{figure*}
	\centering
	\begin{tabular}{cc}
		\epsfxsize=0.5\textwidth
		\epsffile{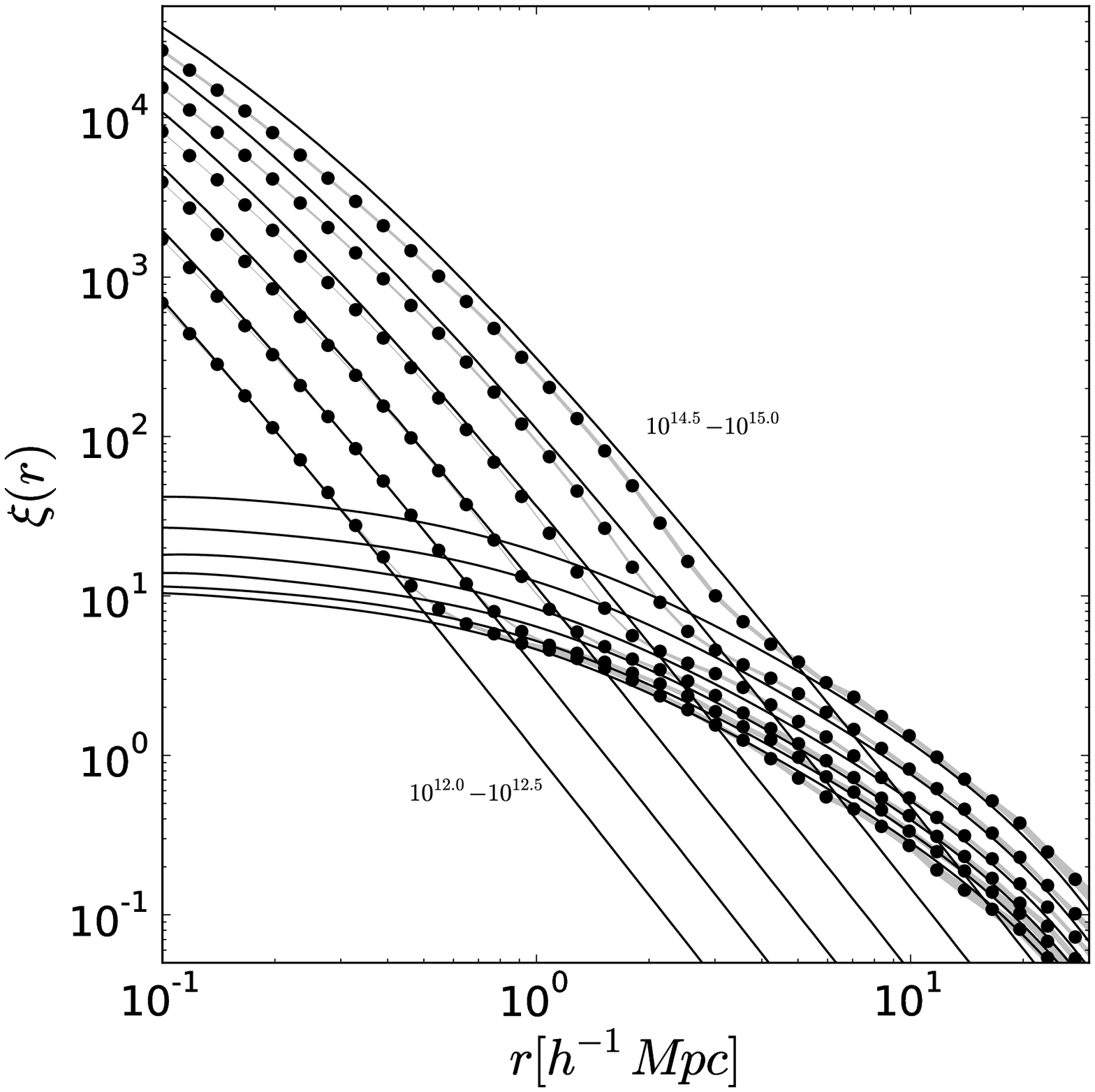} &
		\epsfxsize=0.5\textwidth
		\epsffile{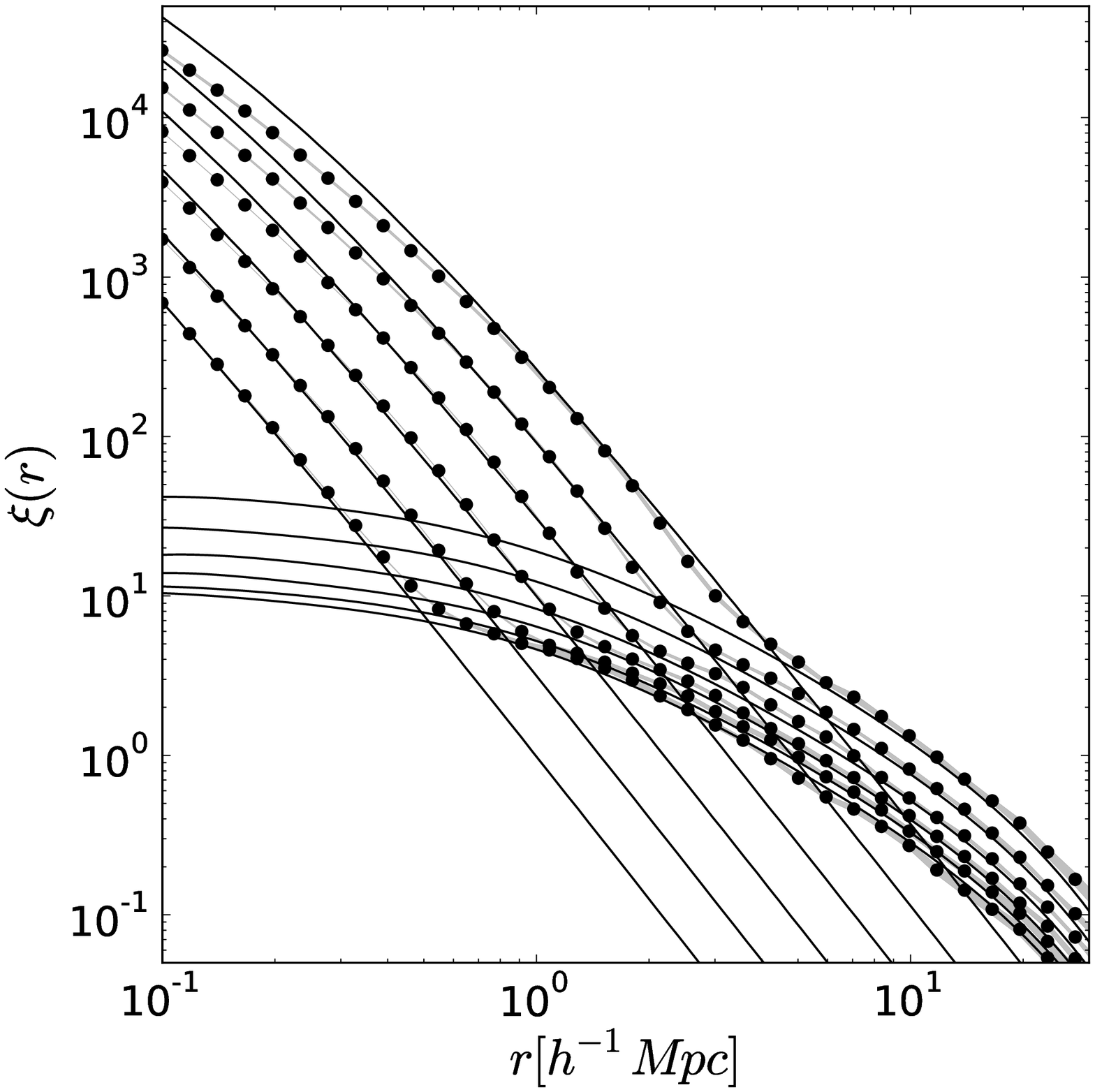}\\
	\end{tabular}
	\caption{Isotropic Halo Model. Filled dots in both panels show the cross correlation functions measured on the
numerical simulation for different mass ranges expanded from $10^{12} M_{\odot}$ to $10^{15} M_{\odot}$. Solid lines
show the results obtained by computing the anisotropic halo model averaging over a conical volume with angle equal to
$\pi$. The 1-halo and 2-halo terms are showed separately. On the left panel the model results are obtained by setting $A
= B = 0.99$ and $C = D = 1.00$, whereas, in the right panel, $A$ and $B$ parameters correspond to the best values
obtained following the $\chi^2$ Minimization Method. As can be appreciated the model obtained using the best values
describes better the numerical results than the model with spherical halo profiles.}
\label{fig:iso}
\end{figure*}

\subsection{Cosmological Simulation and estimation of the properties of the dark matter haloes}
\label{sec:simulation}
In order to test our model, we have used a collisionless simulation of $1024^3$ particles covering a periodic volume of
$500^3$ $(h^{-1} {\rm Mpc})^3$. The initial conditions at redshift $\sim 50$ were calculated assuming a spatially flat
low-density Universe with cosmological parameters taken from the Seven-Year WMAP results \citep{Wmap7} (matter density
$\Omega_{\rm m}=1-\Omega_{\Lambda}=0.272$, Hubble constant $H_{\circ}=70.2$ km s$^{-1}$ Mpc$^{-1}$, and normalization
parameter $\sigma_{8}=0.807$). With these parameters the resulting particle resolution is $m_{\rm p} = 8.78819\times
10^{9}\,h^{-1}\,M_{\odot}$.  The run has been performed using the second version of the GADGET code developed by
\citet{Gadget2}.\\

The identification of particle clumps was carried out by means of a standard friends-of-friends algorithm with a
percolation length of $l=0.17$ $\bar{\nu}^{-1/3}$, where $\bar{\nu}$ is the mean number density of DM particles. For
this study we only kept dark matter haloes with at least $20$ particles.\\

As usual, halo shapes $\vec a$ and orientation $\vec \epsilon$ are defined by means of eigenvalues and eigenvectors of
the shape tensor obtained from halo particle distributions \citep[see for instance][]{Paz2006,Paz2011}. The anisotropic
cross-correlation function is then defined relative to this halo shape axes, by following the same procedure described
in \citet{Paz2011}. This method computes cross-correlation functions by counting halo-particle pairs in conical volumes
around the eigenvector directions. The angles of these volumes are selected such that they are mutually disjoint. In
order to compare these correlation functions with our analytical model, we compute the average anisotropic halo model
over equivalent volumes. To this end, we have computed the correlation functions by averaging our halo model over a
conical section around the three shape axes directions. This average is performed over the two angular components of
$\vec{r}$ on equations \ref{xi1hf} and \ref{xi2hf} increasing the number of integrals by a number of two in each
equation.

At this point, we are ready to compute the anisotropic halo model for a given set of $A$, $B$, $C$ and $D$ parameters.
In the following subsection we estimate this set of parameters through a maximum likelihood method. However, we will
first try to recover the isotropic classic halo model by averaging over a spherical volume (obtained by taking a conical
volume with angle equal to $\pi$). Figure \ref{fig:iso} shows the results for this computation as a function of
centre-halo mass ranging from $10^{12.0} - 10^{15.0} M_{\odot}$ on logarithmic intervals of $0.5$. The 1- and 2-halo
integrals are showed separately. The measurements of the cross correlation function on the numerical simulation
described above are showed as filled circles for the same mass ranges. Jacknife errors are represented by the shaded
grey bands. On the left hand panel, the models are obtained setting $A = B = 0.99$ and $C = D = 1.00$. With this choice
for $A$ and $B$ parameters, dark matter haloes have roughly spherical shapes. The right side panel shows the results
when the parameters $A$ and $B$ are set to their best value, estimated using the minimization method described below. As
can be appreciated the differences between the simulation and both models are more important at the scales corresponding
to the 1-halo term. Additionally, the model with triaxial haloes describes better the numerical results than the model
with spherical haloes.

We have found that the model with triaxial haloes has an average discrepancy of $\%5$ with the results obtained from
numerical simulations, whereas the model with spherical haloes has an average discrepancy of $\%20$. It should be
noticed that the discrepancy between the simulation and the models is more significant for high mass haloes. This is an
expected behaviour since the haloes have more non-spherical profiles. This is in qualitatively agreement with the work
of \citet{SmithWatts} and \citet{2012vanDaalen}. In the first paper \citep{SmithWatts} the authors computed the classic
power spectrum of matter, finding a small but noticeable effect of halo shapes at large frequencies (small scales). On
the other hand, the authors of the second work \citep{2012vanDaalen} have found an effect of $\%20$ on the galaxy
correlation function at small scales when satellite galaxies are sphericalized around the central galaxy. In the same
direction, \citet{SmithWattsSheth2006} found a similar behaviour on the bispectrum. They show that the use of triaxial
haloes in the bispectrum model results in a suppression of $\approx 7\%$ on scales $k > 0.2 h Mpc^{-1}$ relative to a
model based on spherical haloes. Moreover, as our anisotropic model, their approach is insensitive to halo shapes on
large scales.

\subsection{Estimation of parameters: Maximun Likelihood Method}
\label{MLM}

\begin{figure*}
\centering
\begin{tabular}{c}
 \epsfxsize=1.0\textwidth
 \epsffile{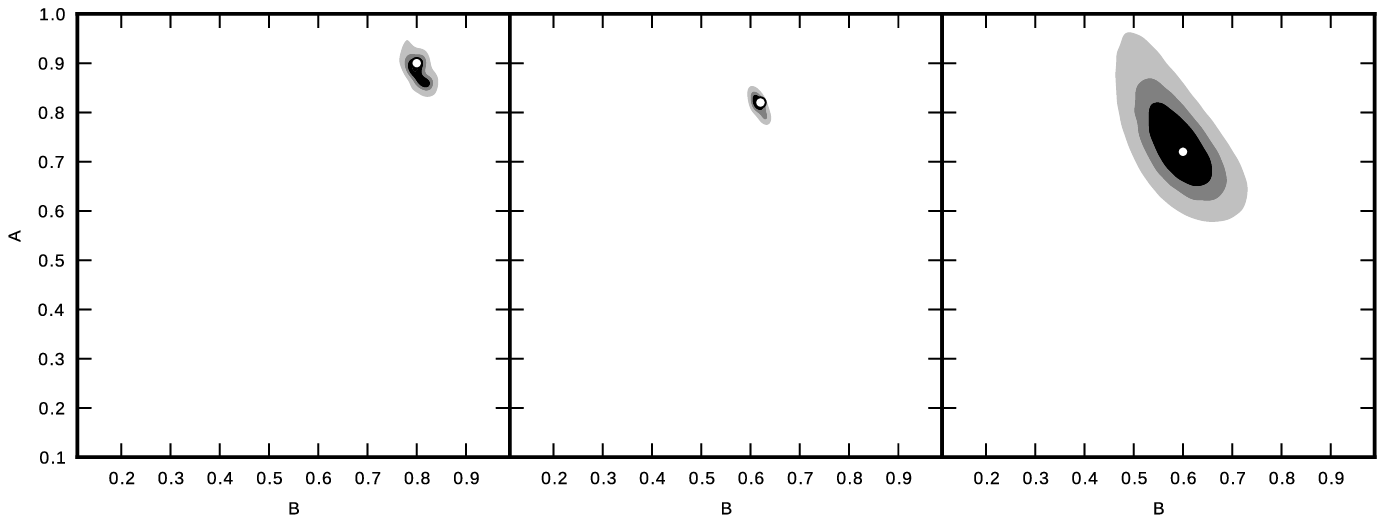}\\
 \epsfxsize=1.0\textwidth
 \epsffile{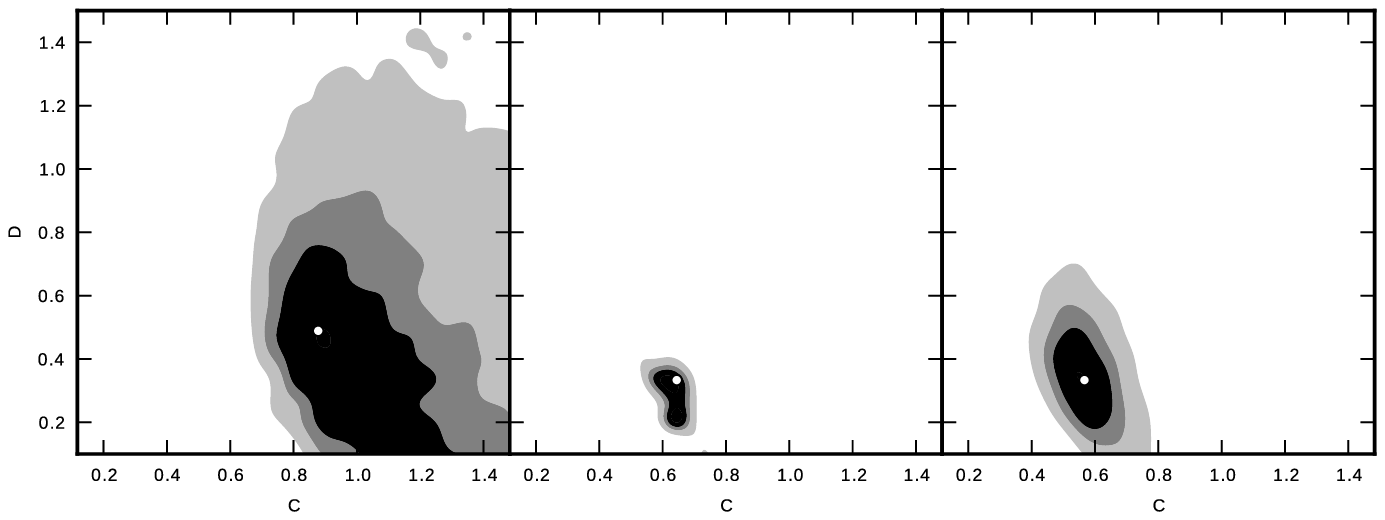}
\end{tabular}
\caption{Likelihood levels of the parameter estimation for three different mass ranges, $10^{12} - 10^{12.5}$,
$10^{13.5} - 10^{14}$ and $10^{14.5} - 10^{15}$, from left to right. Upper panels show the results for $A$ and $B$
parameters while lower panels show the likelihood levels for $C$ and $D$ parameters.}
\label{fig:chis}
\end{figure*}

Before computing and comparing the anisotropic halo model with the simulations, we need to determine the $A$, $B$, $C$
and $D$ parameters. To this purpose we use a standard $\chi^2$ minimization method. Given that the parameters $A$ and
$B$ affect principally the 1-halo term while $C$ and $D$ are only involved in the 2-halo term calculation, we explore
the parameter space with two sets of two parameters, separately. We determine these parameters by fitting, on the one
side, the 1-halo term ($A$ and $B$) and, on the other side, the 2-halo term ($C$ and $D$). This is performed by fitting
each halo term over two disjoint scale intervals separated to avoid mixture between the both terms (see light grey
vertical stride on figure \ref{fig:cono}).

To estimate the best fitting parameters we apply the Maximum Likelihood Estimation method. Figure \ref{fig:chis} shows
the likelihood levels of the parameter estimation for three different mass ranges, $10^{12} - 10^{12.5}$, $10^{13.5} -
10^{14}$ and $10^{14.5} - 10^{15}$, from left to right respectively. Upper panels show the results for $A$ and $B$
parameters while lower panels show the likelihood levels for $C$ and $D$ parameters. The shaded isocontour regions in
black, grey and light-grey colour, show the $1-\sigma$, $2-\sigma$ and $3-\sigma$ levels for the likelihood function,
respectively. The filled white circles on the black region indicates the best fitting set of parameters.

As expected, most massive haloes have a less spherical shape and more prolate mass profile. This can be seen on the
upper panels of the figure \ref{fig:chis} where more massive haloes show lower $A$ and $B$ parameters, meanwhile the
ratio $A/B$ fairly increases. These results are consistent with current agreements about halo shapes and structure
formation \citep{Paz2006,2006Altay_Shape,RobotGroups,2011Lau}. It is worth to mention that the uncertainty in the
parameter space growth with the halo mass. This can be understood in terms of the goodness of our model to fit the
simulations. The adequacy of our model at the 1-halo regime is better for low mass haloes than for high mass haloes. 

The lower panels of figure \ref{fig:chis} show the behaviour of the $C$ and $D$ parameters estimation. It can be seen
that while $C$ decreases with the halo mass, the $D$ parameter increases. On the other hand, the $1-\sigma$ region is
smaller for high than for low mass haloes. The shape of the isocontours on the left panel is elongated along the $D$
parameter. This is due to the weak dependence of the alignment function on the $D$ parameter for large $C$ values.
\subsection{Anisotropic halo model applied to the simulation}

\begin{figure}
	\centering
	\begin{tabular}{c}
		\epsfxsize=0.45\textwidth
		\epsffile{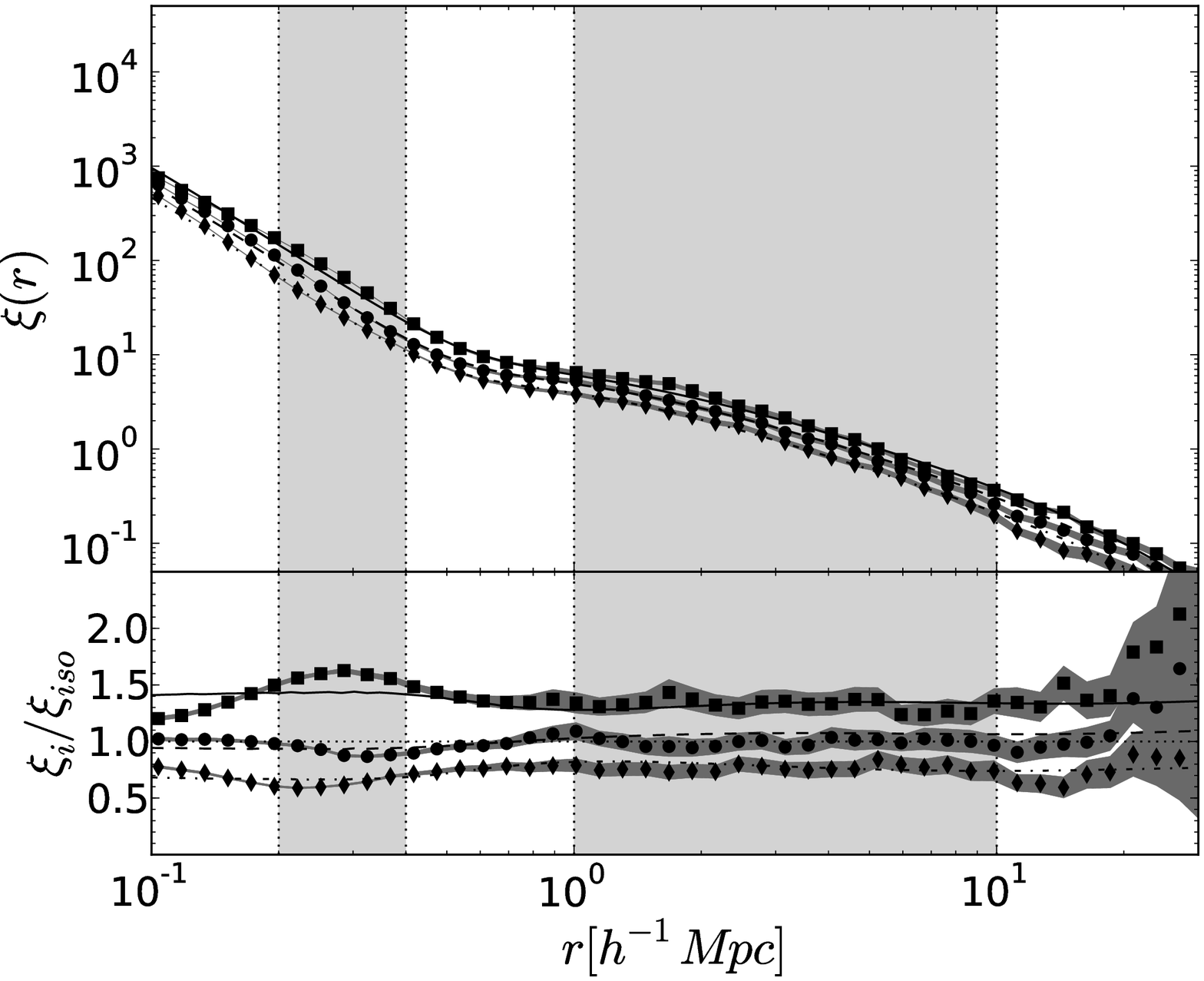} \\
		\epsfxsize=0.45\textwidth
		\epsffile{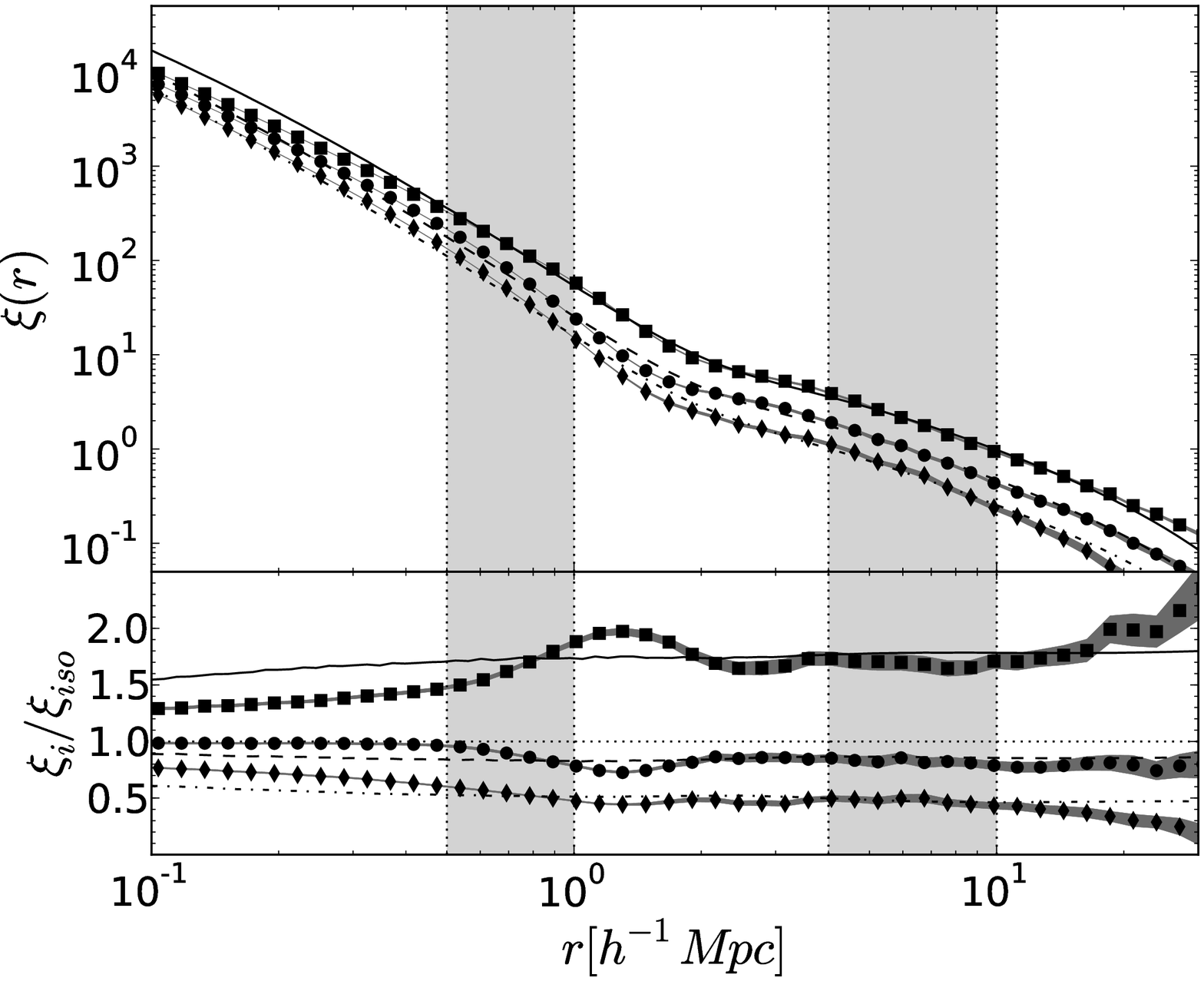} \\
		\epsfxsize=0.45\textwidth
		\epsffile{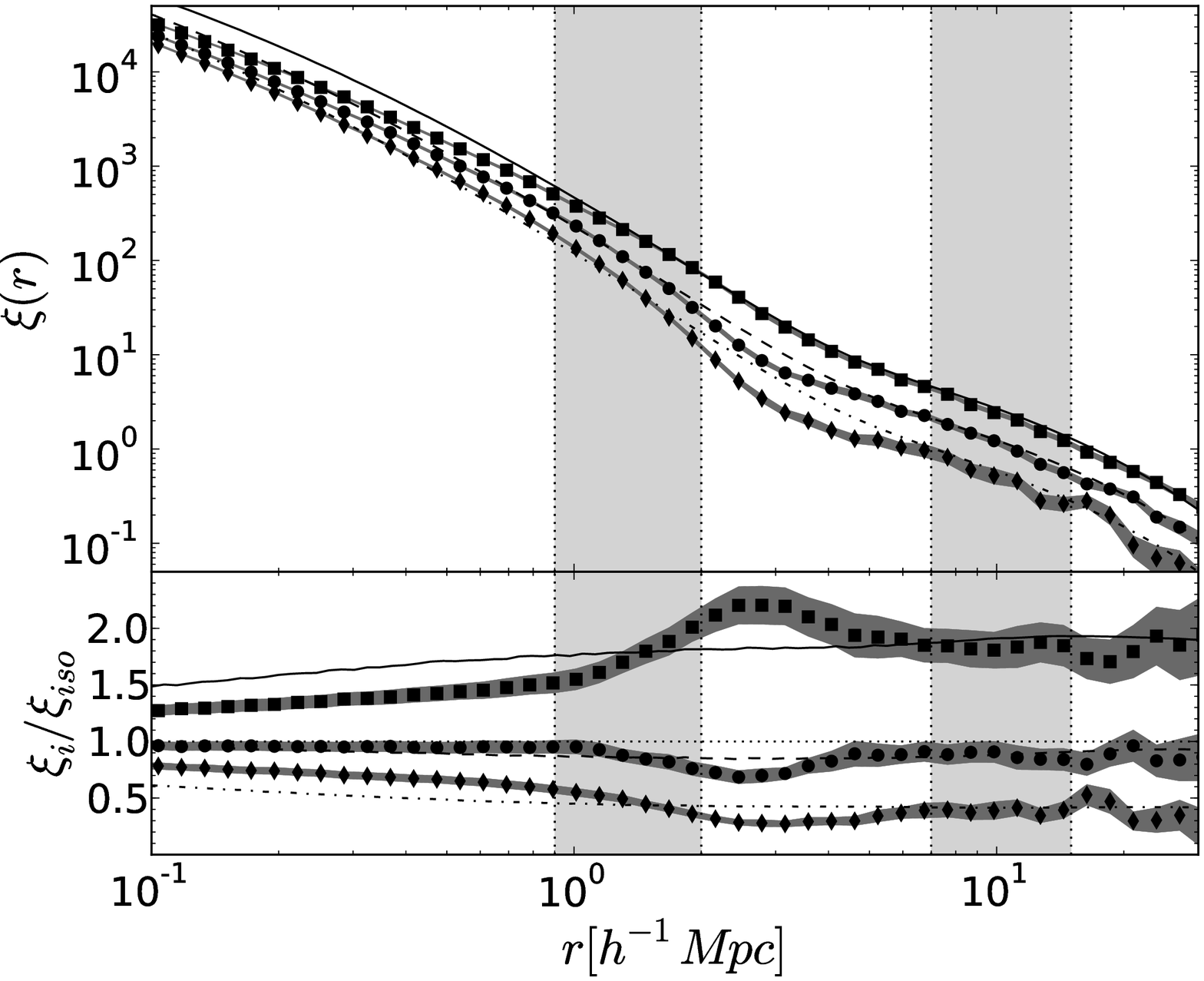}
	\end{tabular}
	\caption{Anisotropic Halo Model for centre haloes in the mass range $10^{12.00}-10^{12.5}$, $10^{13.50}-10^{14.0}$ and
$10^{14.50}-10^{15.0}$ from up to down, respectively. Solid lines show the results obtained by our model with parameters
set to the best value estimated as described in the text. The scale ranges used for these estimations are indicated by
vertical light grey strides. Black points show the measurements of the cross-correlation from numerical simulation with
the corresponding errors in shaded dark grey bands.}
\label{fig:cono}
\end{figure}

Using the best parameter values estimated above we compute the anisotropic halo model for the same three mass ranges
defined in the previous subsection. The results are showed on figure \ref{fig:cono} where each panel corresponds to low,
intermediate and high mass sample from top to bottom. Dotted lines in the upper part of each panel show the anisotropic
cross-correlation functions estimated from the numerical simulation. Each of these measurements are computed adopting a
semi-angle equal to $\pi/4$ for the conical volume. Correlation function measurements along the major, intermediate and
minor shape axes are showed in squares, circles and diamonds, respectively. Dark grey shaded bands around these values
correspond to the Jacknife errors at $1-\sigma$ level. Anisotropic Halo Model results are displayed in solid, dashed,
and dot-dashed lines, corresponding to major, intermediate and minor axes, respectively. At the bottom of each panel, we
show the quotients of each correlation function along shape axis directions to their corresponding isotropic cross
correlation function for the model and the simulation. The light grey vertical bands indicate the scale ranges used to
estimate the best fitting parameters. As can be seen, for the three alignment functions, there is a good agreement
between anisotropic halo model predictions and simulation measurements.

It is worth to mention that the predictions of our model describe better the simulation results in the 2-halo than
in the 1-halo regime. This could be arise from the fact that simulated dark matter profiles become more spherical
towards the centre, whereas our adopted JS profile assumes constant shape. This can be appreciated in lower box of each
panel in figure \ref{fig:cono} where the quotients of simulation measurements become smaller at small scales while for
our models remain roughly constant. On the other hand the predictions of our model describe successfully the simulation
results on the 2-halo regime over a broad scale range. 

\section{Conclusions}
\label{sec:Conclusions}
In this work we have developed a formalism to compute the anisotropic halo-matter cross-correlation function. Main
ingredients added to the classic halo model are the triaxial nature of halo profiles, their probability distribution on
axis ratios and the probability of alignments between two haloes. Since we are interested to quantify the anisotropic
alignment of dark matter haloes through the cross-correlation functions, we have chosen to develop this formalism on
real space opposite to the standard halo model which is computed on Fourier space. This choice impose the calculation of
multidimensional integrals in both, 1-halo and 2-halo terms, which were estimated by implementing a Monte Carlo
integration technique.

As mentioned in the previous paragraph, we have introduced two probability functions. One of them characterising halo
shape distribution and the other describing the alignment between dark matter haloes. Each of these functions has two
free parameters which allow to quantify the influence of halo alignment and halo shape on the anisotropic
cross-correlation function.

As a first test of our algorithm, we have compared the isotropic cross-correlation functions obtained from a numerical
simulation with the results of the anisotropic halo model averaging on a spherical volume. We have found that taking the
triaxial nature of dark matter haloes into account improves at least $\%15$ the predictions of the standard halo model,
as can be seen on figure \ref{fig:iso}. This improvement is more noticeable on scales corresponding to the 1-halo
regime. This is in the same direction as pointed by \citet{SmithWatts} in the sense that triaxial averaging of halo
shapes produce small but noticeably deviation from the classic model. Further comparison in a quantitative way are
difficult to perform since we compute anisotropic cross-correlation functions whereas these authors only calculate the
isotropic power spectrum.

We have compared the predictions of our model with the results obtained from a numerical simulation, and we have
estimated the best fitting parameters of the alignment and shape probability functions by means of a maximum likelihood
method. We have found that our model is able to reproduce the numerical measurements over a wide range of scales,
particularly in the 2-halo regime. Moreover, the model parameters obtained by fitting these numerical results recover,
as expected, the well known mass dependence of halo shapes and the alignment of dark matter haloes with the surrounding
structure. The parameters $C$ and $D$ describe, in a intuitive way, the effect on the cross-correlation function of the
possible alignment configuration between haloes at large scales. On the other side, the parameters $A$ and $B$ reflect
the halo triaxiality at small scales, nevertheless the simplicity of the adopted profile does not allow a fit as good as
in the case of the 2-halo regime. 

In a forthcoming paper we shall consider and develop the presented model in order to compare theoretical predictions
with anisotropic correlation functions measured on galaxy catalogues.

\section*{Acknowledgements}

We would like to thank the anonymous referee for her/his constructive and useful comments and suggestions.
MAS, DP and MM acknowledge support from CONICET and SECyT, Universidad Nacional de C\'ordoba.

\label{lastpage}

\end{document}